%% file: Heurist.tex
\documentstyle[12pt]{article}
\title{Heuristic Models of Two-Fermion Relativistic Systems with 
Field-Type Interaction}

\author{Askold DUVIRYAK\\
Department for Metal and Alloy Theory,\\
Institute for Condensed Matter Physics\\
of Ukrainian National Academy of Sciences,\\ 
1 Svientsitskyj Street, UA-79011, Lviv, Ukraine,\\ 
tel./fax:  (380) 322 761978,\\
e-mail: {\tt duviryak@ph.icmp.lviv.ua}
}

\newcommand{\ha}{\mbox{\small$\frac{1}{2}$}}
\newcommand{\qu}{\mbox{\small$\frac{1}{4}$}}
\newcommand{\se}{\mbox{\small$\frac{3}{2}$}}
\newcommand{\im}{\mathop{\rm i}}
\newcommand{\na}{{\bar a}}
\newcommand{\lab}[1]{\label{#1}}
\newcommand{\re}[1]{(\ref{#1})}

\newcommand{\B}[1]{{\bf#1}}
\newcommand{\Bb}[1]{\mbox{\boldmath$#1$}}
\newcommand{\s}[1]{{\sf#1}}
\newcommand{\G}[1]{\Gamma_{\rm#1}}
\newcommand{\Si}[1]{\Sigma_{\rm#1}}
\newcommand{\bs}{{\Bb\sigma}}
\newcommand{\V}[1]{V_{\rm#1}}
\newcommand{\W}[1]{W_{\rm#1}}
\newcommand{\bra}{\langle}
\newcommand{\ket}{\rangle}
\newcommand{\bbra}{\mbox{\boldmath$\langle$}}
\newcommand{\bket}{\mbox{\boldmath$\rangle$}}

\begin{document}
\maketitle
\begin{abstract}
	We use the chain of simple heuristic expedients to obtain 
perturbative and exactly solvable relativistic spectra for a family of 
two-fermionic bound systems with Coulomb-like interaction. In the case of 
electromagnetic interaction the spectrum coincides up to the second order 
in a coupling constant with that following from the quantum electrodynamics. 
Discrepancy occurs only for S-states which is the well-known difficulty 
in the bound-state problem. The confinement interaction is 
considered too.
\vspace{1ex}\\
PACS number(s): 03.65.Pm, 03.65.Ge, 12.39.Pn
\end{abstract}


\section{Introduction}

	A nonrelativistic two-body problem reduces completely to the 
single-body problem with a central potential. The single-body Schr\"odinger 
equation becomes the two-body one (for the centre-of-mass frame of 
reference) if to replace the particle mass by the reduced mass, and 
to understand the radius-vector $\B r$ as the relative position vector. 

	In the relativistic case the relation between the single- and 
two-particle problems is not so transparent. There are few reasons for that.
First, a spin appears on a physical stage. The existence of spin 
diversifies properties of both interacting particles of matter and fields 
mediating this interaction. Thus even in the single-particle problem various 
relativistic wave equations such as the Klein-Gordon, Dirac, Duffin-Kemmer 
and other equations exist, and they can involve scalar, vector, tensor and 
other interaction potentials. Second, there exist different approaches to 
the relativistic two- and few- body problem. The most profound approaches 
are based on the quantum field theory (QFT), especially on the perturbative 
QFT. They lead to complicated integral equations (such as the Bethe-Salpeter 
equation \cite{Sal52,BS57}), coupled sets of differential equations (such as 
the Breit equation \cite{Breit29,BS57}), or their higher order 
differential reductions (such as the Fermi-Pauli equation 
\cite{BLP71,BS57}). Other approaches such as quasipotential 
\cite{Tod71}--\cite{RST85} or ones based on the relativistic direct 
interaction theory (RDIT) \cite{Hav71}--\cite{GTY98} are semi- or purely 
phenomenological. They manifest a general structure of relativistic 
potentials and wave equations while features of concrete interactions must 
be brought from other sources (for example, from the classical or quantum 
field theory). 

	I.~Todorov has observed a simple way how the Klein-Gordon equation 
can be transformed into the quasipotential equation describing the scalar 
and vector interaction of two spinless particles \cite{Tod71,RST85}. In the 
case of Coulomb-like interaction this equation gives the spectrum which 
agrees with QFT result up to $\alpha^4$--terms of coupling constant 
expansion. The generalization for an arbitrary field-type interaction 
(including higher rank tensor interactions) arises naturally from the 
Schwinger source theory \cite{Sch70} and the Fokker formalism 
\cite{Hav71}, and results in some RDIT models
\cite{Duv97}--\cite{D-S01}. In the present paper we construct relativistic 
wave equations appropriate for the description of the field-type interaction 
of two particles with spin $\ha$.

	At the beginning we summarize results concerning spinless system. 
Namely, in Section 2 we formulate the family of relativistic wave equations 
which describe the scalar, vector and gravitational (i.e., tensor) 
interaction of two scalar particles. These equations have a common 
effective single-particle structure. In the case of Coulomb-like 
interaction they are exactly solvable. The mass spectra coincide with that 
obtained algebraically in Ref. \cite{D-S01}, and agree up to $\alpha^4$ 
with known QFT results. The only disagreement exists for S-states.

	Then we modify the wave equations in order to describe two-fermion 
systems. The spin-orbital, spin-spin and tensor corrections to the scalar, 
vector and gravitational interactions are brought from QFT\footnote{We 
note that two-particle spin-orbital interaction can be recovered completely 
from the single-particle one \cite{C-O'C77} but this is not concerned with 
spin-spin and tensor terms.} (Section 3). In Section 4 by means of an 
appropriate rescaling of $\B r$ spin interaction is included into the 
eigenstate problem as small correction to the effective Coulomb 
Hamiltonian. Calculations with the first order perturbation theory 
(Section 5) reproduce the QED muonium spectrum up to 
$\alpha^4$ \cite{Her92} and give a generalization for the cases of 
scalar and gravitational interactions.

	Spin potential corrections depend on $r=|\B r|$ as $1/r^3$. Due to 
properties of matrix elements with Coulomb bound states we transform
spin corrections into $1/r^2$ terms in such a way that 
perturbative spectrum (up to $\alpha^4$) does not change. In this form the 
problem appears exactly solvable, which is shown explicitly in Section 6.
	
	The Todorov construction was proposed in the case of Coulomb-like 
interactions. Here (in Section 7) we approve this recipe for a system with 
confining interaction, and make an appropriate modification to
account spin effects.


\section{Spectra of systems of two spinless particles}

	Let us consider the stationary Klein-Gordon equation for particle 
of the rest mass $m$ in the scalar potential $\V{s}(r)$ and the vector 
one $\V{v}(r)$ depending on $r=|\B r|$:
%
\begin{equation}\lab{2.1}
\Delta\Psi(\B r) + \left([E - \V{v}(r)]^2 - [m + \V{
s}(r)]^2\right)\Psi(\B r) = 0.
\end{equation}
Following the Todorov's observation in the quasipotential approach 
\cite{Tod71,RST85} one can construct the appropriate two-particle wave 
equation by the following substitution:
%
\begin{equation}\lab{2.2}
E \to E_M = \frac{M^2 - m_1^2 - m_2^2}{2M}, \qquad
m \to m_M = \frac{m_1m_2}{M},
\end{equation}
where $m_a$ is the rest mass of $a$th particle, and $M$ is the total mass of 
the system, i.e., the energy in the centre-of-mass frame of reference. One 
obtains the Schr\"odinger-like equation:
%
\begin{equation}\lab{2.3}
\Delta\Psi(\B r) + [Q_M - U_M(r)]\Psi(\B r) = 0,
\end{equation}
where
%
\begin{equation}\lab{2.4}
Q_M \equiv E_M^2 - m_M^2 = \frac{1}{4M^2}[M^2 - m_+^2][M^2 - m_-^2],
\end{equation}
is the on-shell value of the relative momentum squared as a function of $M$
\cite{RST85},
%
\begin{equation}\lab{2.5}
U_M(r) = 2[m_M \V{s}(r) + E_M \V{v}(r)] + 
\V{s}^2(r) - \V{v}^2(r)
\end{equation}
is the effective potential, and $m_\pm=m_1\pm m_2$. In the nonrelativistic 
limit the equation \re{2.3} becomes the usual Schr\"odinger equation with 
the nonrelativistic potential $V(r)=\V{s}(r) + \V{v}(r)$ where $r$ is the 
distance between particles.

	The effective potential $U_M(r)$ depends of the total mass $M$. 
Thus the equation \re{2.3} is of the quasipotential type, and 
problems can occur with the consistent quantum-mechanical treatment 
\cite{RST85}. 

	In the case of Coulomb-like interaction
%
\begin{eqnarray}\lab{2.6}
\V{s}(r)= - \alpha_{\rm s}/r,\\
\lab{2.7}
\V{v}(r)= - \alpha_{\rm v}/r
\end{eqnarray}
with the coupling constants $\alpha_{\rm s}>0$ and $\alpha_{\rm v}>0$ the 
equation \re{2.3} is exactly solvable. Moreover, for the case of 
electromagnetic (vector) interaction the corresponding spectra coincide 
(except for the ground S-states) with those following from the scalar 
quantum electrodynamics in the second order of a coupling constant 
\cite{BIZ70,Tod71}. 

	The simple Todorov recipe embraces the cases of scalar and vector 
interactions (and their superposition). The generalization to the cases of 
higher-rank tensor interactions can be built on the base of 
the family of RDIT models (known as time-asymmetric) 
\cite{Duv97,Duv98}. This leads to the Schr\"odinger-like 
(quasipotential) equation \re{2.3} with
%
\begin{equation}\lab{2.8}
U_M(r) = - 2m_Mf(\lambda)\frac{\alpha}{r} + \gamma\frac{\alpha^2}{r^2},
\end{equation}
where $\lambda=E_M/m_M$, and the function $f(\lambda)$ (such that $f(1)=1$) 
as well as the constant $\gamma$ depend on the tensor nature of interaction 
\cite{Duv98,D-S01}. For example, for the scalar, vector and gravitational 
interaction (or another second-rank tensor interaction, for example, 
the strong gravitation) we have:
%
\begin{eqnarray}
\lab{2.9}
f_{\rm s}(\lambda) &=& 1,\qquad\qquad\quad \gamma_{\rm s}=1,\\
\lab{2.10}
f_{\rm v}(\lambda) &=& \lambda,\qquad\qquad\quad \gamma_{\rm v}=-1,\\
\lab{2.11}
f_{\rm g}(\lambda) &=& 2\lambda^2-1,\qquad\ \gamma_{\rm g}=-6.
\end{eqnarray}
The superposition of these interactions can be considered by means of the 
superposition of the functions and constants \re{2.9}-\re{2.11}.
Also we note that for scalar and vector interactions (and their 
superposition) the effective potentials \re{2.8} (from the time-asymmetric 
models) is identical to \re{2.5} with \re{2.6} or/and \re{2.7} (from the 
Todorov recipe).

	The mass spectrum following from the equation \re{2.3}, \re{2.4}, 
\re{2.8} can be presented in the implicit form:
%
\begin{equation}\lab{2.12}
M^2 = m_1^2 + m_2^2 + 2m_1m_2\lambda,
\end{equation}
where $\lambda$ is a positive solution of the equation:
%
\begin{equation}\lab{2.12a}
\frac{1-\lambda^2}{f^2(\lambda)} = \frac{\alpha^2}{\nu^2},
\end{equation}
and
%
\begin{equation}\lab{2.12b}
\nu = n_r + \ha +\sqrt{(\ell+\ha)^2 + \alpha^2\gamma}
\end{equation}
is effective ``principal quantum number''; here $n_r = 0,1,...$ is the 
radial quantum number and $\ell = 0,1,...$ is the angular quantum number.

	For the scalar, vector and gravitational interactions the equation 
\re{2.12a} is solvable:
%
\begin{eqnarray}\lab{2.12c}
\lambda_{\rm s} &=& \sqrt{1 -\alpha^2/\nu^2},
\\ \lab{2.12d}
\lambda_{\rm v} &=& 1\left/\sqrt{1 +\alpha^2/\nu^2}\right.,
\\ \lab{2.12e}
\lambda_{\rm g} &=& \frac1{2\sqrt{2}} \sqrt{4 - \frac{\nu^2}{\alpha^2} + 
\frac{\nu}{\alpha}\sqrt{8 + \nu^2/\alpha^2}}.
\end{eqnarray}

	Approximately, with accuracy up to $\alpha^4$, we have :
%
\begin{equation}\lab{2.13}
M \approx m_+ - \frac{m_r\alpha^2}{2n^2} + \frac{m_r\alpha^4}{2n^4}
\left[f^\prime(1)-\frac14 -\frac{m_r}{4m_+}\right] + \frac{m_r\alpha^4}{n^3}
\frac{\gamma}{2\ell+1},
\end{equation}
where $f^\prime=df/d\lambda$, $m_r=m_1m_2/m_+$ is the reduced mass, and 
$n=n_r+\ell+1$ is the principal quantum number.


\section{Two-fermion systems: including spin effects}

	The weakly relativistic system of two fermions interacting via scalar 
or/and vector field can be described by the generalized Breit-Fermi 
Hamiltonian \cite{O-M83,L-S89}. Besides the nonrelativistic Coulomb 
Hamiltonian, it includes relativistic kinematic terms, spin-independent and 
spin-dependent corrections to the interaction. Some of these terms are 
singular and can be taken into account as perturbations only.

	Here we do not consider the Breit-Fermi Hamiltonian.
Instead, we modify the Todorov recipe in order to describe the spin 
effects in two-fermionic systems. For this purpose we need only a
spin-dependent part $\W{}$ of the Breit-Fermi Hamiltonian. For the scalar 
and vector interaction it is \cite{O-M83,L-S89}:
%
\begin{eqnarray}
\lab{3.1}
\W{s} &=& -\frac14\,\B L\cdot\left(\frac{\bs_1}{m_1^2}+
\frac{\bs_2}{m_2^2}\right)\frac{\V{s}^\prime(r)}{r},\\
\lab{3.2}
\W{v} &=&\ \frac14\,\B L\cdot\left(
\left[\frac1{m_1^2}+\frac2{m_1m_2}\right]\bs_1 +
\left[\frac1{m_2^2}+\frac2{m_1m_2}\right]\bs_2\right)
\frac{\V{v}^\prime(r)}{r}\nonumber\\
&&{}+\frac1{12m_1m_2}\left(\frac1r \V{v}^\prime(r)-
\V{v}^{\prime\prime}(r)\right)T +
\frac1{6m_1m_2}\,\bs_1\cdot\bs_2\,\Delta
\V{v}(r),
\end{eqnarray}
where $\B L=-\im\B r\times\Bb\nabla$ is the orbital momentum operator,   
$\bs_a$ is the spin operator acting on the $a$th particle 
spin variable, and $T=3(\bs_1\cdot\B n)(\bs_2\cdot\B n)-\bs_1\cdot\bs_2$ is 
the tensor operator. In the case of gravitational interaction (with the
nonrelativistic potential $\V{g}(r)= - \alpha_{\rm g}/r$, where 
$\alpha_{\rm g}=Gm_1m_2$, and $G$ is the gravitational constant) we have 
\cite{BGH66}:
%
\begin{eqnarray}
\lab{3.3}
\W{g} &=&\ \frac14\,\B L\cdot\left(
\left[\frac3{m_1^2}+\frac4{m_1m_2}\right]\bs_1 +
\left[\frac3{m_2^2}+\frac4{m_1m_2}\right]\bs_2\right)
\frac{\alpha_{\rm g}}{r^3}\nonumber\\
&&{}+\frac{\alpha_{\rm g}}{4m_1m_2r^3}\,T +
\frac{2\pi\alpha_{\rm g}}{3m_1m_2}\,\bs_1\cdot\bs_2\,\delta(\B r).
\end{eqnarray}

	Now in order to construct a two-fermion equation we replace a 
nonrelativistic potential $V$ by $\tilde V=V+\W{}$ in the effective 
potential $U_M$ (Eq. \re{2.5} or \re{2.8}). The resulting quasipotential 
equation is not solvable, and we should apply some approximate 
method.

	In the case of Coulomb-like interaction the spin term $\W{}$ is 
meant to be small as to compare to the nonrelativistic potential 
$V(r)=-\alpha/r$. Thus we can modify approximately the effective potential 
\re{2.8} as follows:
%
\begin{equation} \lab{3.4}
U_M(r) \to \tilde U_M(r) \approx U_M(r)  + 2m_Mf(\lambda)\W{}.
\end{equation}

	Now one can account the spin correction by means of the perturbation 
method. In so doing we note the following. First, the original 
(non-perturbed) equation \re{2.3} fails to describe correctly S-states. Thus 
we will neglect $\delta$-functional terms in $\W{}$ (i.e., the last term in 
r.h.s. of Eq. \re{3.3}, and the last term in r.h.s. of Eq. \re{3.2} in the 
case \re{2.7}) as they contribute in S-states only. Second, the modified
equation (as well as the unperturbed one \re{2.3}) is quasipotential but not 
the true Schr\"odinger equation. Thus it needs some minor reformulation 
to be tractable within the perturbation method.


\section{Spin corrections to the Co\-u\-lomb-like interaction}

	In the case of Coulomb-like interaction the two-fermion wave equation
reads:
%
\begin{equation} \lab{4.1}
\left[\B p^2 - 2m_Mf(\lambda)\left(\frac{\alpha}{r} - W\right) 
+ \gamma\frac{\alpha^2}{r^2} - Q_M\right]\Psi = 0,
\end{equation}
where $\B p = -\im\Bb\nabla$. Using the substitution
%
\begin{equation} \lab{4.2}
\B r = \Bb\rho/R_M,\qquad \B p = R_M\Bb\pi\qquad {\rm with}\quad 
R_M = \alpha m_M 
f(\lambda)
\end{equation}
we present the equation \re{4.1} in the dimensionless 
Hamiltonian form:
%
\begin{equation} \lab{4.3}
H\Psi = \varepsilon\Psi.
\end{equation}
Here
%
\begin{equation} \lab{4.4}
H=H^{(0)} + H^{(1)} 
\end{equation}
is the total Hamiltonian,
%
\begin{equation} \lab{4.5}
H^{(0)} = \frac12\Bb\pi^2 - \frac1\rho
\end{equation}
is the basic Coulomb Hamiltonian, 
%
\begin{equation} \lab{4.6}
H^{(1)}=\alpha^2\G{},\qquad \G{} = \frac{\gamma}{2\rho^2} + 
\frac1{\rho^3}\,\Si{}(\B n)
\end{equation}
is a perturbation since $\alpha$ is considered as a 
small parameter, and
%
\begin{equation} \lab{4.7}
\varepsilon = \frac{Q_M}{2R_M^2} = \frac{\lambda^2-1}{2\alpha^2f^2(\lambda)}
\end{equation}
is a dimensionless energy (i.e., a spectral parameter).

	With sufficient accuracy (i.e., up to terms $\sim O(\alpha)$) the 
last term of $\G{}$ is equal to $W/(\alpha^3R_M)$ but does not depend on $M$.
This is provided by use of approximate equality $M\approx m_+$ in small 
terms. The general form of the operator $\Si{}$ acting on angular and spin 
variables is:
%
\begin{equation}\lab{4.8}
\Si{} = \left[(\xi-\delta^2)\B L\cdot\bs_+ +
2\eta\delta\B L\cdot\bs_- + \zeta(1-\delta^2)T\right]/16,
\end{equation}
where $\bs_\pm=\bs_1\pm\bs_2$, $\delta=m_-/m_+$, and the constants 
$\xi$, $\eta$ and $\zeta$ 
for scalar, vector and gravitational interactions are defined as follows:
%
\begin{eqnarray}
\lab{4.9}
\xi_{\rm s} &=& -1,\qquad\quad \eta_{\rm s}=1,\ \qquad\quad 
\zeta_{\rm s}=0,\\
\lab{4.10}
\xi_{\rm v} &=& 3,\ \qquad\quad \eta_{\rm v}=-1,\qquad\quad 
\zeta_{\rm v}=1,\\
\lab{4.11}
\xi_{\rm g} &=& 7,\ \qquad\quad \eta_{\rm v}=-3,\qquad\quad 
\zeta_{\rm v}=1.
\end{eqnarray}
%


\section{Basic states and first-order perturbation theory}

	The basic Hamiltonian \re{4.5} commutes with
operators of orbital angular momentum $\B L$, total spin $\B 
S=\ha\bs_+$, total angular momentum $\B J = \B L + \ha\bs_+$ and parity 
$P$. In order to write down the basic eigenfunctions $\Psi^{(0)}(\Bb\rho)$ 
we use the angular ``bispinor harmonics'' $\varphi^i(\B n)$\
($i=A,0,-,+$). In 2$\times$2 matrix representation they are \cite{D-L96}:
%
\begin{equation} \label{5.1}
\varphi^A({\B n})= {1\over {\sqrt {2}}}Y^\mu_j({\B n})
\left[  
\begin{array}{cc}
0&-1\\ 1&0
\end{array}
 \right],
\end{equation}
\begin{equation} \label{5.2}
\varphi^0({\B n}) = {1\over {\sqrt {2{j}({j}{+}{1})}}}
\left[
\begin{array}{ccc}
-\sqrt{({j}{-}{\mu}{+}{1})({j}{+}{\mu})}\,Y^{\mu-1}_{J}&&{\mu}\,Y^{\mu}_{j}\\
&&\\
{\mu}\,Y^{\mu}_{j}&&\sqrt{({j}{+}{\mu}{+}{1})({j}-{\mu})}\,Y^{\mu+1}_{j}  
\end{array}
\right],
\end{equation}
\begin{equation} \label{5.3} 
\varphi^-({\B n}) = {1\over {\sqrt {2({j}{+}{1})({2}{j}{+}{3})}}}  
\left[
\begin{array}{cc}
\sqrt{({j}{-}{\mu}{+}{1})({j}{-}{\mu}{+}{2})}Y^{\mu-1}_{j+1}&
-\sqrt{({j}{+}{\mu}{+}{1})({j}{-}{\mu}{+}{1})}Y^{\mu}_{j+1}\\
&\\
-\sqrt{({j}{+}{\mu}{+}{1})({j}{-}{\mu}{+}{1})}Y^{\mu}_{j+1}&
\sqrt{({j}{+}{\mu}{+}{1})({j}{+}{\mu}{+}{2})}Y^{\mu+1}_{j+1}
\end{array}
\right],  
\nonumber
\end{equation}
\begin{equation} \label{5.4}
\varphi^+({\B n}) = {1\over {\sqrt {{2}{j}({2}{j}{-}{1})}}}
\left[
\begin{array}{ccc}
\sqrt{({j}{+}{\mu}{-}{1})({j}{+}{\mu})}\,Y^{\mu-1}_{j-1}&&
\sqrt{({j}{+}{\mu})({j}{-}{\mu})}\,Y^{\mu}_{j-1}\\
&&\\
\sqrt{({J}{+}{\mu})({j}{-}{\mu})}\,Y^{\mu}_{j-1}&&
\sqrt{({j}{-}{\mu}{-}{1})({j}{-}{\mu})}\,Y^{\mu+1}_{j-1}
\end{array}
\right],
\end{equation}
where $Y_\ell^\mu(\B n)$\ ($\mu = -\ell,...,\ell$) are the spherical 
harmonics depending on the direction $\B n=\B r/r$.
The bispinor harmonics form an orthonormal set, in the sense that 
$\bra i | k \ket =\int d\B n \, {\rm Tr}(\varphi_{i}^{\dag} \, 
\varphi_{k})=  \delta_{i\,k}$, where the integrations 
are taken over the entire solid angle. 

	The bispinor harmonics posses the following properties (besides 
those due to properties of the spherical harmonics):
%
\begin{equation}\lab{5.5}
\begin{array}{rclcrcl}
\B L\cdot\bs\varphi^A &=&\sqrt{j(j+1)}\varphi^0, &&
\B L\cdot\bs\varphi^- &=& -(j+2)\varphi^-, \\
\B L\cdot\bs\varphi^0 &=&\sqrt{j(j+1)}\varphi^A -\varphi^0, &&
\B L\cdot\bs\varphi^+ &=& (j-1)\varphi^+, \\
\B n\cdot\bs\varphi^{A,0} &=& -\sqrt{\frac{j+1}{2j+1}}\varphi^\mp 
\pm \sqrt{\frac{j}{2j+1}}\varphi^\pm, &&
\B n\cdot\bs\varphi^\mp &=& -\sqrt{\frac{j+1}{2j+1}}\varphi^{A,0} 
\mp \sqrt{\frac{j}{2j+1}}\lefteqn{\varphi^{0,A},}
\end{array}
\end{equation}
where the components of the vector operator $\bs$ are the Pauli matrices.

	The action of spin operators on the bispinor harmonics is as follows: 
$\bs_1\varphi=\bs\varphi$, $\bs_2\varphi=\varphi\bs^T$.
We note that $\varphi^A$ is antisymmetric and $\varphi^{0,\pm}$ are
symmetric matrices. Then it follows from this and Eqs. \re{5.5} that 
$\varphi^A$ and $\varphi^{0,\pm}$ satisfy the following equalities:
%
\begin{equation}\lab{5.6}
\begin{array}{rclcrcl}
\B J^2\varphi &=& j(j+1)\varphi, &&
j &=&0,1,..., \\
J_3\varphi &=& \mu\varphi, &&
\mu &=&-j,...,j, \vspace{.25ex}\\
\B L^2\varphi^i &=& \ell(\ell+1)\varphi^i, &&
\ell &=&\left\{{\displaystyle\begin{array}{lcl}
j, && i=A,0,\\
j\pm1, && i=\mp,
\end{array}}\right.\vspace{.25ex}\\
\B S^2\varphi^i &=& s(s+1)\varphi^i, &&
s &=& \left\{{\displaystyle\begin{array}{lcl}
0, && i=A,\\
1, && i=0,\mp,
\end{array}}\right.\vspace{.25ex}\\
P\varphi^{A,0} &=& (-)^j\varphi^{A,0}, &&
P\varphi^\mp &=& (-)^{j+1}\varphi^\mp, \\
\ha\B L\cdot\bs_+\varphi^A &=&0, &&
\ha\B L\cdot\bs_+\varphi^- &=& -(j+2)\varphi^-, \\
\ha\B L\cdot\bs_+\varphi^0 &=& -\varphi^0, &&
\ha\B L\cdot\bs_+\varphi^+ &=& (j-1)\varphi^+, \\
\ha\B L\cdot\bs_-\varphi^{A,0} &=& \sqrt{j(j+1)}\varphi^{0,A}, &&
\ha\B L\cdot\bs_-\varphi^\mp &=& 0, \\
\bs_1\cdot\bs_2\varphi^A &=& -3\varphi^A, &&
\bs_1\cdot\bs_2\varphi^{0,\mp} &=& \varphi^{0,\mp}, \\
\ha T\varphi^A &=& 0, &&
\ha T\varphi^- &=&3\,\frac{\sqrt{j(j+1)}}{2j+1}\varphi^+
-\frac{j+2}{2j+1}\varphi^- , \\
\ha T\varphi^0 &=& \varphi^0, &&
\ha T\varphi^+ &=&3\,\frac{\sqrt{j(j+1)}}{2j+1}\varphi^-
-\frac{j-1}{2j+1}\varphi^+ .
\end{array}
\end{equation}

	Now one can choose four independent basic eigenfunctions 
$\Psi_i^{(0)}(\Bb\rho)$ ($i=A,0,-,+$) of $H^{(0)}$ as follows:
%
\begin{eqnarray}
\lab{5.7}
\Psi_{A,0}^{(0)}(\Bb\rho) = \frac1\rho u_{n,j}(\rho)\varphi^{A,0}(\B n),&&
\Psi_\mp^{(0)}(\Bb\rho) = \frac1\rho u_{n,j\pm1}(\rho)\varphi^\mp(\B n),
\end{eqnarray}
where $u_{n,\ell}(\rho)$ is a solution of the radial Coulomb problem 
%
\begin{equation}\lab{5.8}
H_{\ell}u_{n,\ell}(\rho) = \varepsilon^{(0)}u_{n,\ell}(\rho)
\end{equation}
with the effective Hamiltonian
%
\begin{equation} \lab{5.9}
H_\ell = -\frac12\left\{\frac{d}{d\rho^2} -  
\frac{\ell(\ell+1)}{\rho^2}\right\} - \frac{1}{\rho}
\end{equation}
and the dimensionless eigenenergy
%
\begin{equation}\lab{5.10}
\varepsilon^{(0)}=-1/(2n^2),\quad n=1,2,\dots
\end{equation}
We note that the basic eigenfunctions $\Psi_{A,0}^{(0)}(\Bb\rho)$ 
have the parity $P=(-)^j$, and $\Psi_\mp^{(0)}(\Bb\rho)$ have the parity 
$P=(-)^{j+1}$. The function $\Psi_A^{(0)}(\Bb\rho)$ describes the singlet
($s=0$, $\ell=j$) state while $\Psi_{0,\mp}^{(0)}(\Bb\rho)$ correspond to 
triplet ($s=1$, $\ell=j,j\pm1$) states.

	Let us calculate the first-order correction $\varepsilon^{(1)}$ to
the dimensionless energy $\varepsilon \approx \varepsilon^{(0)} + 
\alpha^2\varepsilon^{(1)}$. The total Hamiltonian $H=H^{(0)}+\alpha^2\G{}$ 
commutes with operators of parity $P$ and total angular momentum $\B J = \B 
L + \ha\bs_+$. One can choose the wave functions $\Psi(\Bb\rho)$ as the 
eigenfunctions of $\B J^2$, $J_3$ and $P$. Thus they can be spanned onto 
states $\Psi_{A,0}^{(0)}$ if $P=(-)^j$, or onto $\Psi_\mp^{(0)}$ if 
$P=(-)^{j\mp1}$. In the each parity case zero-order eigenvalues 
$\varepsilon^{(0)}$ are twice degenerated. Thus in the first order of 
perturbation theory we have
%
\begin{equation}\lab{5.11}
\varepsilon^{(1)}_{(i,k)}=\frac12\left[{\G{}}_{ii}+{\G{}}_{kk}\pm
\sqrt{({\G{}}_{ii}-{\G{}}_{kk})^2 + 4{\G{}}_{ik}^{\,2}}\right]\qquad (i\ne 
k)
\end{equation}
with $i=A,~k=0$ if $P=(-)^j$ and $i=-,~k=+$ if $P=(-)^{j+1}$, where the 
matrix $\s\Gamma=\left[{\G{}}_{ik}\right]$ is defined as follows:
%
\begin{equation}\lab{5.12}
\s\Gamma=\left[{\G{}}_{ik}\right] = \left[\bbra i|\G{}| k\bket\right] =\left[
\int d\B r\, {\rm Tr}\left(\Psi^{\dag}_i(\B r)\G{}\Psi_k(\B r)\right)\right].
\end{equation}
Taking \re{4.6} and \re{5.7} into account we have:
%
\begin{equation}\lab{5.13}
\s\Gamma = \frac{\gamma}{2}\bra j|\rho^{-2}|j\ket\s1 + \bra j|\rho^{-3}|j\ket
\left[\begin{array}{ccc}
\bra A|\Si{}|A\ket && \bra A|\Si{}|0\ket \\
&&\\
\bra 0|\Si{}|A\ket && \bra 0|\Si{}|0\ket 
\end{array}\right]
\end{equation}
if $P=(-)^j$, and
%
\begin{eqnarray}\lab{5.14}
\s\Gamma &=& \frac{\gamma}{2}
\left[\begin{array}{ccc}
\bra {j}{+}{1}|\rho^{-2}|{j}{+}{1}\ket && 0 \\
&&\\
0 && \bra {j}{-}{1}|\rho^{-2}|{j}{-}{1}\ket 
\end{array}\right] \nonumber\\
&&+ 
\left[\begin{array}{ccc}
\bra {j}{+}{1}|\rho^{-3}|{j}{+}{1}\ket \bra A|\Si{}|A\ket &&
\bra {j}{+}{1}|\rho^{-3}|{j}{-}{1}\ket  \bra A|\Si{}|0\ket \\
&&\\
\bra {j}{-}{1}|\rho^{-3}|{j}{+}{1}\ket \bra 0|\Si{}|A\ket &&
\bra {j}{-}{1}|\rho^{-3}|{j}{-}{1}\ket \bra 0|\Si{}|0\ket 
\end{array}\right]
\end{eqnarray}
if $P=(-)^{j+1}$, where
%
\begin{equation}\lab{5.15}
\bra i|\Si{}|k\ket =\int d\B n\, {\rm Tr}\left(\varphi^{\dag}_i(\B 
n)\Si{}\varphi_k(\B n)\right)
\end{equation}
and
\begin{equation}\lab{5.15a}
\bra \ell^\prime|\rho^s|\ell\ket = \int d\rho\,u_{n,\ell^\prime}(\rho)\,
\rho^s\, u_{n,\ell}(\rho).
\end{equation}
In particular,
	
%
\begin{equation}\lab{5.16}
\bra \ell|\rho^{-2}|\ell\ket = \frac{1}{n^3(\ell+\ha)},\qquad
\bra \ell|\rho^{-3}|\ell\ket = 
\frac{\bra \ell|\rho^{-2}|\ell\ket}{\ell(\ell+1)},
\end{equation}
\begin{equation}\lab{5.17}
\bra {\ell}{+}{1}|\rho^{-2}|{\ell}{-}{1}\ket = 0,\qquad
\bra {\ell}{+}{1}|\rho^{-3}|{\ell}{-}{1}\ket = 0.
\end{equation}
The relations \re{5.16} are well known in literature (see \cite{L-L58} or
\cite{Boh86}), and \re{5.17} can be calculated by means of 
formulae given in \cite[chap. {\em Mathematical Supplements}, $\S$~f]{L-L58}.

	Using \re{5.16}, \re{5.17} and calculating the matrix elements $\bra 
i|\Si{}|k\ket$ by means of Eqs.
\re{4.8}--\re{4.11}, \re{5.6} one obtains the matrix $\s\Gamma$ 
and then the corrections $\varepsilon^{(0)}$ to the dimensionless energy.
Then, using \re{2.12}, \re{4.7} and expanding the total mass $M$ in 
$\alpha$ one obtains the first-order mass spectra (i.e., with accuracy up to 
$\alpha^4$).
	
	Due to the relations \re{5.16} and \re{5.17} the matrix $\s\Gamma$
is not diagonal if $P=(-)^j$. Thus the correspondent first-order states are 
the mixture of singlet ($s=0$, $\ell=j$) and triplet ($s=1$, $\ell=j$) 
states. In the $P=(-)^{j+1}$ case $\s\Gamma$ is diagonal, and the triplet 
($s=1$, $\ell=j\pm1$) states does not mix. Thus it is convenient to 
classify the first-order mass spectra by $j$ and $\ell$. These spectra can 
be obtained from Eq. \re{2.13} by the following substitution:
%
\begin{equation}\lab{5.18}
\gamma \to \gamma + \phi(\ell,j)
\end{equation}
where the function $\phi(\ell,j)$ depends on both a spin state of the system 
and the tensor rank of mediating field. We have:
%
\begin{eqnarray}
\lab{5.19}
\phi_{\rm s} &=& \left\{
\begin{array}{lcl}
\displaystyle{
\frac1{8\ell(\ell+1)}\left(1+\delta^2\pm
\sqrt{(1+\delta^2)^2 + 16\delta^2\ell(\ell+1)}\right),
}&& \ell=j,
\vspace{1ex}\\
\displaystyle{
\frac{1+\delta^2}{4\ell},
}&& \ell=j+1,
\vspace{1ex}\\
\displaystyle{
-\frac{1+\delta^2}{4(\ell+1)},
}&& \ell=j-1,
\end{array}\right. \\ \nonumber\\
\lab{5.20}
\phi_{\rm v} &=& \left\{
\begin{array}{lcl}
\displaystyle{
-\frac1{4\ell(\ell+1)}\left(1\pm
\sqrt{1 + 4\delta^2\ell(\ell+1)}\right),
}&& \ell=j,
\vspace{1ex}\\
\displaystyle{
-\frac1{2\ell}-
\frac{1-\delta^2}{2(2\ell-1)},
}&& \ell=j+1,
\vspace{1ex}\\
\displaystyle{
\frac1{2(\ell+1)}+
\frac{1-\delta^2}{2(2\ell+3)},
}&& \ell=j-1,
\end{array}\right. \\ \nonumber\\
\lab{5.21}
\phi_{\rm g} &=& \left\{
\begin{array}{lcl}
\displaystyle{
-\frac3{4\ell(\ell+1)}\left(1\pm
\sqrt{1 + 4\delta^2\ell(\ell+1)}\right),
}&& \ell=j,
\vspace{1ex}\\
\displaystyle{
-\frac3{2\ell}-
\frac{1-\delta^2}{2(2\ell-1)},
}&& \ell=j+1,
\vspace{1ex}\\
\displaystyle{
\frac3{2(\ell+1)}+
\frac{1-\delta^2}{2(2\ell+3)},
}&& \ell=j-1.
\end{array}\right. 
\end{eqnarray}

	The Eqs. \re{2.13}, \re{5.18} and \re{5.20} reproduce the muonium 
spectrum \cite{Her92} and (if $m_1=m_2=m$) the positronium spectrum 
\cite{BLP71}.

\input Heurist1.tex

\input Heuristl.tex

\end{document}

%% file: Heurist1.tex

\section{Solvable simulation of first-order mass spectra}

	Solving the Schr\"odinger equation \re{4.3} perturbatively is due to 
the fact that spin interaction term in the operator \re{4.6} depends 
on $\rho$ as $\rho^{-3}$. Below we construct some exactly solvable model 
which reproduces the spectrum of perturbation theory.

	Let us modify the operator \re{4.6} as follows
%
\begin{equation}\lab{6.1}
\G{} \longrightarrow \tilde\Gamma  = \frac{Z(\B n)}{2\rho^2},
\end{equation}
where 
%
\begin{equation}\lab{6.2}
Z(\B n) = \gamma + 2\{\Si{}(\B n)/\B L^2\}_{\rm ordered}.
\end{equation}
The operator $Z$ acts on angle and spin variables. It is not defined on 
states which contain the S-wave, but we refuse these states from the very 
beginning. On other states $Z$ is supposed to be Hermitian. Thus it must be 
somehow ordered if $\Si{}(\B n)$ and $\B L^2$ do not commute.  

	It is easy to examine by means of Eqs. \re{5.16}--\re{5.17} that 
in the first order of perturbation theory the Hamiltonian $\tilde H = 
H^{(0)} + \alpha^2\tilde{G}$ has the same spectrum as the original 
Hamiltonian $H$. This result does not depend on the ordering rule used in 
$Z$.

	Below we show that the new Schr\"odinger equation is exactly solvable. 
Of course, the exact solution and corresponding spectrum depend on the 
ordering rule. One can consider, for example, the following rules:
%
\begin{eqnarray}\lab{6.2a}
\{\Si{}/\B L^2\}_{\rm ordered}&=&
\ha(\Si{}|\B L|^{-2} +|\B L|^{-2}\Si{}),
\\\lab{6.2b}
\{\Si{}/\B L^2\}_{\rm ordered}&=&
|\B L|^{-1}\Si{}|\B L|^{-1}
\\\lab{6.2c}
\{\Si{}/\B L^2\}_{\rm ordered}&=&
\int\limits_0^\infty dt\, e^{-\frac t2\B L^2}\Si{}e^{-\frac t2\B L^2},
\end{eqnarray}
where $|\B L|=\sqrt{\B L^2}$. The last rule is inspired by the Feynman 
representation of an inverse operator: $A^{-1}=\int_0^\infty dt\exp(-tA)$.

	The radial reduction of the Schr\"odinger equation can be performed 
by the following choice of the wave functions $\Psi(\Bb\rho)$ as the 
eigenfunctions of $\B J^2$, $J_3$ and $P$:
%
\begin{equation} \lab{6.3}
\Psi(\Bb\rho) = \frac1\rho\sum_i\psi_i(\rho)\varphi^i(\B n).
\end{equation}
Here the summa in r.h.s. of Eq. \re{6.3} runs over 
$i=A,0$ if $P=(-)^j$, and over $i=-,+$ if $P=(-)^{j+1}$. Substituting this
function into the new Schr\"odinger equation and collecting coefficients 
at bispinor harmonics $\varphi^A$ and $\varphi^0$ (or at $\varphi^-$ and 
$\varphi^+$) one obtains the pair of coupled Rarita-Schwinger equations. In 
the matrix form they are:
%
\begin{equation} \lab{6.4}
\s H\s\Psi(\rho) = \varepsilon\s\Psi(\rho),
\end{equation}
where 
%
\begin{equation} \lab{6.5}
\s\Psi(\rho) = \left[\psi_i(\rho)\right]
\end{equation}
is two-component column wave function,
%
\begin{equation} \lab{6.6}
\s H = -\frac12\left\{\frac{d}{d\rho^2} -  
\frac1{\rho^2}\s K\right\} - \frac{1}{\rho},
\end{equation}
and
%
\begin{equation} \lab{6.7}
\s K = \left[K_{ik}\right] = \left[\bra i|\B L^2+\alpha^2 Z|k\ket\right].
\end{equation}

	The form of 2$\times$2 symmetric matrix $\s K$ depends on both the 
parity and the tensor structure of interaction:
%
\begin{eqnarray}
\lab{6.7a}
\!{\s K}\!&{=}&
\!\left[\!\mbox{\scriptsize$
\begin{array}{ccc}
j(j+1) + \alpha^2\gamma &&
\frac{\alpha^2\eta\delta}{2\sqrt{j(j+1)}} \\
&&\\
\frac{\alpha^2\eta\delta}{2\sqrt{j(j+1)}} &&
j(j+1) + 
\alpha^2\left(\gamma-\frac{\xi-\delta^2-\zeta(1-\delta^2)}{4j(j+1)}\right)
\end{array}$}
\!\right]
\end{eqnarray}
for the parity
$P=(-)^j$, and
\begin{eqnarray}
\lab{6.7b}
\!{\s K}\!&{=}&
\!\left[\!\mbox{\scriptsize$
\begin{array}{cc}
(j{+}1)(j{+}2) +
\alpha^2\!\left(\!\gamma-\frac1{4(j+1)}\left[\xi-\delta^2+\zeta
\frac{1-\delta^2}{2j+1}\!\right]\right) 
&
\alpha^2\zeta\frac{1-\delta^2}{4(2j+1)}\frac{3\sqrt{j(j+1)}}{j(j+1)+1} \\
&\\
\alpha^2\zeta\frac{1-\delta^2}{4(2j+1)}\frac{3\sqrt{j(j+1)}}{j(j+1)+1} &
(j{-}1)j +
\alpha^2\!\left(\!\gamma+\frac1{4j}\left[\xi-\delta^2-\zeta
\frac{1-\delta^2}{2j+1}\!\right]\right) 
\end{array}$}
\!\right]
\end{eqnarray}
for the parity $P=(-)^{j+1}$.
We note that in general case where $\zeta\ne0$ (including the cases of 
vector and gravitational interaction; c.f. Eqs.\re{4.10}, \re{4.11}) 
the operators $\Si{}$ and $\B L^2$ does not commute. Thus in 
calculating of $\s K$ we chosen the ordering rule \re{6.2c}.
The use of \re{6.2a} or \re{6.2b} leads to off-diagonal elements of $\s K$ 
which are singular at $j=1$ (besides of $j=0$). But there is no any 
physical reason for such a singularity.

	Using now an appropriate unitary (even orthogonal, to be sharp) 
transformation, the matrix $\s K$ can be diagonalized, so that the coupled 
Rarita-Schwinger equations split into a pair of one-dimensional 
Schr\"odinger equations with effective Hamiltonians 
$H_{\tilde\ell_{(i,k)}}$ of the form \re{5.9} but with non-integer 
$\tilde\ell_{(i,k)}$:
%
\begin{equation}\lab{6.8}
\tilde\ell_{(i,k)} = - \ha + \sqrt{\qu + K_{(i,k)}},
\end{equation}
with 
%
\begin{equation}\lab{6.9}
K_{(i,k)}=\frac12\left[K_{ii}+K_{kk}\pm
\sqrt{(K_{ii}-K_{kk})^2 + 4K_{ik}^{\,2}}\right]\qquad (i\ne k),
\end{equation}
where $i=A,k=0$ if $P=(-)^{j}$, and $i=-,k=+$ if $P=(-)^{j+1}$.

	These equations are exactly solvable and lead to the mass spectrum
\re{2.12}--\re{2.12e} but with another effective ``principal quantum 
number'':
%
\begin{equation}\lab{6.10}
\nu \longrightarrow \tilde\nu = \tilde\nu(n_r,j,P) = 
n_r+\tilde\ell_{(i,k)}+1.
\end{equation}
The calculation of effective ``principal quantum number'' 
$\tilde\nu$ is straightforward by the use of Eqs. \re{6.7a}--\re{6.10}.
Here we do not write down these rather cumbersome formulae.


\section{The confinement problem}

	The Todorov recipe was observed on the systems with Coulomb-like 
interaction. Here we demonstrate that this rule appears useful for 
the construction of relativistic potential model of mesons. 

	It is well known that the spectra of heavy quarkoniums are described 
satisfactory (and modulo spin effects) by means of the nonrelativistic 
potential model with the short-range Coulomb potential \re{2.7} and the 
long-range linear potential \cite{L-S89}:
%
\begin{equation}\lab{7.14}
\V{v}(r)= ar,
\end{equation}
where $a>0$ is a constant. 

	The description of light mesons needs the application of 
relativistic models. They frequently are built as single-particle wave 
equations with the vector short-range potential and the scalar long-range 
one \cite{Ono82}. Other models treat mesons as an extended 
objects or a composite two-quark relativistic systems. One of them which is 
concise and elegant, and which reflects principal features of the light 
meson spectroscopy, is the covariant oscillator model. Few versions of this 
model are given in Refs. \cite{Tak79}.

	The nonrelativistic, single-particle and oscillator models appears 
related to one another by the Todorov recipe. Given the single-particle 
Klein-Gordon equation \re{2.1} with the scalar potential \re{7.14} and the 
vector one \re{2.7}, this rule fixes unambiguously the form of the 
two-particle wave equation \re{2.3}--\re{2.5} which is the relativization 
of the nonrelativistic potential model.

	If $m_a=0$ and $\alpha=0$ the wave equation reduces to the 
oscillator problem and yields the exact solution for the mass spectrum:
%
\begin{equation}\lab{7.15}
M^2 = 8a\left[\ell + 2n_r + \se\right].
\end{equation}
The spectrum falls on the family of straight lines in the 
($M^2$,$\ell$)--plane known in the hadron spectroscopy as the leading 
(for $n_r=0$) and daughter's ($n_r>0$) Regge trajectories. This structure 
and ($\ell+2n_r$)--degeneracy of the spectrum \re{7.15} are characteristic 
of actual light meson spectra, if to neglect the rest mass contribution and 
fine spin effects.

	In the general case the equation \re{2.3}--\re{2.5}, \re{2.7}, 
\re{7.14} is not exactly solvable. Here we use the oscillator approximation 
to estimate the spectrum for $\ell$ large.

	The substitution $\Psi(\B r) = \frac1r\psi(r)Y_\ell^\mu(\B n)$
reduces the equation \re{2.3}--\re{2.5}, \re{7.14} to the form
%
\begin{equation}\lab{7.16}
\psi^{\prime\prime}(r) + [Q_M - U_{M\ell}(r)]\psi(r) = 0,
\end{equation}
where
%
\begin{equation}\lab{7.17}
U_{M\ell}(r) = U_M(r) + \ell(\ell+1)/r^2.
\end{equation}
The function $U_{M\ell}(r)$ has a local minimum at some point 
$r_0$ depending on $M$, $\ell$ and satisfying the condition
%
\begin{equation} \lab{7.18}
U_{M\ell}^{\prime}(r_0)=0.
\end{equation}
Thus one can expand the potential \re{7.17} at the minimum,
%
\begin{equation} \lab{7.19}
U_{M\ell}(r) \approx U_{M\ell}(r_0)  + \frac12 U_{M\ell}^{\prime\prime}(r_0) 
(r-r_0)^2,
\end{equation}
and search a solution of this oscillator problem. A quantization condition 
then reads:
%
\begin{equation} \lab{7.20}
Q_M - U_{M\ell}(r_0) = \sqrt{\ha U_{M\ell}^{\prime\prime}(r_0)}(2n_r+1).
\end{equation}
If $n_r\sim1$, the approximate solution differs exponentially little from
the exact solutions of the problem.

	Eqs. \re{7.18} and \re{7.20} form the set of algebraic equations with 
$r_0$ and $M$ to be found. Solving this set by a power series in $\ell$ 
leads to the asymptotic formulae:
%
\begin{equation} \lab{7.21}
r_0^2 = \frac{\ell + \ha}{a} - \frac1{\sqrt2a}\left(\frac{m_1m_2}{4a} +
\alpha\right) +  O(\ell^{-1})
\end{equation}
and
%
\begin{equation} \lab{7.22}
M^2 = 8a\left[\ell + 2n_r + \se - \sqrt2\alpha\right] +
2\left(m_1^2 + m_2^2 + \sqrt2m_1m_2\right) +  O(\ell^{-1}).
\end{equation}
The latter represents the spectrum of the system at $\ell\gg n_r$ (but 
provides a good approximation even if $\ell\simeq2\div3$).
As to compare this formula to the Eq. \re{7.15}, the influence of the 
short-range interaction and non-zero rest masses result in the parallel 
shift of the family of Regge trajectories as a whole. Considering the 
constants $m_1$, $m_2$ (and, possibly, $\alpha$) as adjustable parameters 
one can obtain trajectories for different families of light mesons. 

	Up to now we neglected a mass splitting due to a spin interaction.
The majority of attempts to describe spin effects in hadron 
spectroscopy concerns with the heavy quark systems which can be treated as 
weakly relativistic systems. As usual, one takes the generalized Pauli-Fermi 
Hamiltonian with long-range scalar potential and short-range vector one 
(the potentials \re{7.14}, \re{2.7} in our case) and corresponding spin 
corrections \re{3.1}, \re{3.2} treated perturbatively 
\cite{O-M83,L-S89,HIK82}. But this scheme can fail 
when considering light mesons as corresponding to strongly relativistic 
domain $M\gg m_{1(2)}$. First of all we note that, as it follows from Eqs. 
\re{7.21}, \re{7.22}, in this domain $r_0^2\sim\ell$ and $M^2\sim\ell$. 
Thus $r_0\sim M$, i.e., the radius of meson is proportional to its mass; 
here we took into account that the wave function at $\ell$ large is 
localized around $r_0$. Then we have rough estimates:
%
\begin{eqnarray}
\V{s}=ar\sim M,\qquad &\W{s}\sim\ell/r\sim M \quad &\Longrightarrow \quad
\V{s}\sim \W{s},
\nonumber\\\lab{7.23}
\V{v}=-\alpha/r\sim M^{-1},\quad &\W{v}\sim\ell/r^3\sim M^{-1}
\quad &\Longrightarrow \quad\V{v}\sim \W{v}.
\end{eqnarray}
The spin corrections appears to be of the same order as the nonrelativistic 
potentials. But actual spin effects in light meson spectra are small. 
Moreover, the operators \re{3.1} and \re{3.2} was deduced within the 
perturbation theory \cite{O-M83,L-S89}, so they should satisfy inequality 
$W\ll V$ by construction. Second, the operators  \re{3.1} and 
\re{3.2} are divergent if $m_1$~or/and~$m_2\to0$. Consequently, divergent 
terms appear in the mass spectrum (in contrast to the case of Coulomb-like 
interaction where the spectrum is not singular if $m_1$~or/and~$m_2$ 
vanish). 

	The possible way to avoid these two problems is the following 
modification of operators \re{3.1} and \re{3.2}: the rest masses of quarks 
involved in these operators should be replaced by the ``constituent'' 
masses:
%
\begin{equation} \lab{7.24}
m_a \to M_a = \sqrt{m_a^2 + Q_M} = 
\frac{M^2 + m_a^2-m_{\na}^2}{2M},\qquad a=1,2,\quad \na=3-a
\end{equation}
possessing the properties 1)~$M_1+M_2=M$, 2)~$M_a\approx M/2$ if $M\gg 
m_{1(2)}$, and 3)~$M_a\to m_a$ in the nonrelativistic limit $M\to m_+$ (we 
suppose that $M^2>|m_1^2-m_2^2|$). The modified operators $\overline W$ are 
regular at $m_1$~or/and~$m_2\to0$, they became small as $\overline W\sim 
V(m_+/M)^2$ in strongly relativistic domain $M\gg m_{1(2)}$, and reduce to 
$W$ in weakly relativistic domain $M\approx m_+$. 

	Now let us estimate the effect of spin corrections $\overline\W{s}$ 
to the scalar potential \re{7.14} (the effect of vector interaction is 
minor, as it follows from the estimates in \re{7.23} and the paragraph 
above). For this purpose we modify the scalar potential 
$\overline\V{s}=\V{s}+ \overline\W{s}$ and substitute it (instead of 
$\V{s}$) into the effective potential $U_M$. Then we diagonalize the 
modified effective potential $\overline U_M(r)$ and apply the oscillator 
approximation to the resulting pair of split quasipotential equations. 
Finally, we come to the asymptotically linear Regge trajectories \re{7.22} 
with the same slope parameter $8a$, but each trajectory splits into three 
ones by parallel shift 
%
\begin{equation}\lab{7.26}
\Delta\!M^2 =\left\{
\begin{array}{lcl}
\displaystyle{
\ 0,
}&& \ell=j,
\vspace{1ex}\\
\displaystyle{
\pm2a,
}&& \ell=j\pm1.
\end{array}\right. 
\end{equation}
Note that the mass splitting does not depend on $\ell$. Qualitatively this 
result correlates with actual light meson spectra as well as with 
theoretical results following from the string models 
\cite{PPS87,BNP92}.  

	As in the spinless case, the model is exactly 
solvable if $m_a=0$ and $\alpha=0$. The mass spectrum is determined by 
$M^2=4ax$, where $x$ is the positive solution of the transcendental 
equation:
%
\begin{equation} \lab{7.27}
x^2 - 2(2n_r+1)x - \kappa = \sqrt{(2\ell+1)^2x^2+\kappa^2}.
\end{equation}
which, in turns, reduces to a cubic algebraic equation, and
%
\begin{equation}\lab{7.28}
\kappa =\left\{
\begin{array}{lcl}
\displaystyle{
0,
}&& \ell=j,\ s=0,
\vspace{1ex}\\
\displaystyle{
1,
}&& \ell=j,\ s=1,
\vspace{1ex}\\
\displaystyle{
\ell+1,
}&& \ell=j+1,\ s=1,
\vspace{1ex}\\
\displaystyle{
-\ell,
}&& \ell=j-1,\ s=1.
\end{array}\right. 
\end{equation}
Here we do not write down an explicit form of the mass spectrum.

	Of course, the minimal set of adjustable parameters makes the 
present potential model too poor to provide a sharp fit to an experimental 
data. In particular, the mass splitting \re{7.26} is unambiguously fixed by 
the typical hadron scale $8a$. This value (0\% and 25\% of $8a$) 
contradicts to some estimates of actual experimental data
(about 5$\div$6\%; see \cite{BNP92}). 

	In another version of the model we include operators $\overline W$ 
similarly to the case of Coulomb-like interaction (see Eq. \re{3.4}), i.e., 
by the following modification of the effective potential:
%
\begin{equation} \lab{7.29}
U_M(r) \to \overline U_M(r) = U_M(r)  + 2m_M\overline W_{\rm s} +
2E_M\overline W_{\rm v}.
\end{equation}
In the strongly relativistic domain the present potential model leads to 
the same asymptotic spectrum \re{7.22} as the spinless model does.
The mass splitting is small. It can be calculated perturbatively and 
fitted to experimental data by adjusting the parameters $m_a$ and $\alpha$.


\section{Summary}

	In the present paper we have constructed the family of simple 
quantum-mechanical models which describe two-fermion relativistic systems
with different Coulomb-like and confining interactions. We have embodied in 
these models some theoretical experience of studies in the relativistic 
two-body problem. We started with the quasipotential equations describing 
the scalar and vector interaction of two spinless particles. As it was 
observed by Todorov \cite{Tod71,RST85}, these equations are solvable, 
and they have simple single-particle structure. Similar features are 
characteristic of the family of RDIT models \cite{Duv98,D-S01} 
which describe an arbitrary relativistic Coulomb-like interaction including 
the gravitation. We generalize these equations to the case of two-fermionic 
systems. For this purpose we use the operators known from QFT which
describe spin-dependent corrections to the scalar, vector and 
gravitational interacting \cite{BLP71,O-M83,L-S89,BGH66}. We first treat 
these operators perturbatively and obtain the spectrum of muonium and its 
scalar and gravitational counterparts with accuracy up to $\alpha^4$.
Then we modify the equations in such a way that they become exactly 
solvable and yield correct (within the same accuracy) mass spectra.
These equations can be useful for accounting higher than $\alpha^4$ (say, 
radiative) corrections to a particle interaction by means of the first 
order perturbation theory.

	Also we have demonstrated that the Todorov recipe of constructing 
two-body equations permits a straightforward application to the case of 
confining interaction. The generalization to spinning particles has 
been proposed too. In the weakly relativistic domain this equation reduces 
to well-known potential model \cite{L-S89} which is appropriate to 
the description of heavy mesons. In the strongly relativistic limit it 
yields a mass spectrum which reproduces qualitatively light-meson 
experimental data.

%% file: Heuristl.tex

%% file: Heurist.bbl
\begin{thebibliography}{99}
\bibitem{Sal52}
E. E. Salpeter, Phys. Rev. {\bf 87}, 328 (1952).
\bibitem{BS57}
H. A. Bethe and E. E. Salpeter, {\sl Quantum mechanics of
One- and Two-Electron Atoms} (Springer, Berlin, 1957).
\bibitem{Breit29}
G. Breit, Phys. Rev. {\bf 34}, 553 (1929).
\bibitem{BLP71}
V. B. Berestetskii, E. M. Lifshitz and L. P. Pitaevskii, {\em Relativistic
quantum theory}, (Pergamon Press, Oxford, 1st Edition, 1971).
\bibitem{Tod71}
I. T. Todorov, Phys. Rev. D {\bf 3}, 2351 (1971).
\bibitem{BIZ70}
E. Brezin, C. Itzykson and J. Zinn-Justin, Phys. Rev. D {\bf 1},
2349 (1970).
\bibitem{RST85}
V. A. Rizov, H. Sazdjian and I. T. Todorov, Ann. Phys. (NY) {\bf 165}, 59
(1985).
\bibitem{Hav71}
P. Havas, ``Galilei- and Lorentz-invariant 
particle  systems  and their conservation laws'',
in  {\sl Problems  in  the  Foundations  of Physics},
M. Bunge, ed.,
Berlin, Springer, 1971, pp.~31--48;
P. Ramond, Phys. Rev.~D {\bf 7}, 449 (1973);
V. I. Tretyak, {\em Fokker-Type Action Integrals and Forms of
Relativistic Lagrangian Dynamics}. Thesis,
Doctor of Science, Lviv State University, Lviv (1996).
\bibitem{GTY98}
R. Gaida, V. Tretyak and Yu. Yaremko, Cond. Matter Phys. {\bf 1}, 425
(1998).
\bibitem{Sch70}
J. Schwinger, {\em Particles, Sources, and Fields}, (Addison-Willey, Pub.
Co., Massachusetts, 1970).
\bibitem{Duv97}
A. Duviryak, Acta Phys. Polon. {\bf B28}, 1087 (1997).
\bibitem{Duv98}
A. Duviryak, Gen. Rel. Grav. {\bf 30}, 1147 (1998);
A. Duviryak, V. Shpytko and V. Tretyak, Cond. Matter Phys. {\bf 1}, 463
(1998).
\bibitem{D-S01}
A. Duviryak and V. Shpytko, Rep. Math. Phys. {\bf 48}, 219 (2001).
\bibitem{C-O'C77}
L.-H. Chan and R. F. O'Connell, Phys. Rev. D {\bf 15}, 3058 (1977).
\bibitem{Her92}
J. H. Connell, Phys. Rev. D {\bf 43}, 1393 (1991); H. Hersbach, Phys. Rev.
A {\bf 46}, 3657 (1992).
\bibitem{O-M83}
M. G. Olsson and K. J. Miller, Phys. Rev. {\bf D28}, 674 (1983); R. W.
Childers, Phys. Rev. {\bf D36}, 606, 3813 (1987).
\bibitem{L-S89}
W. Lucha and F. F. Sch\"oberl, {\em Die Starke Wechselwirkung. Eine
Einf\"uhrung in nichtrelativistische Potentialmodelle} (Bibliographishches
Institut \& F. A. Brockhaus, Mannheim, 1989).
\bibitem{BGH66}
B. M. Barker, S. N. Gupta and R. D. Haracz, Phys. Rev. {\bf 149}, 1027
(1966).
\bibitem{D-L96}
Ju. Darewych and L. Di Le, J. Phys. A, {\bf 29}, 6817 (1996).
\bibitem{L-L58}
L. D. Landau and E. M. Lifshitz, {\em Quantum Mechanics: non-relativistic
theory}, (Pergamon Press, London, 1st Edition, 1958).
\bibitem{Boh86}
A. Bohm, {\em Quantum Mechanics: Foundations and Applications},
(Springer-Verlag, New York, 1986).
\bibitem{Ono82}
S. Ono, Phys. Rev. D {\bf 26}, 2510 (1982);
C. Goebel, D. LaCourse and M. G. Olsson, Phys. Rev. D {\bf 41}, 2917
(1990); I. Haysak, V. Lengyel, A. Shpenik, S. Challupka and M. Salak, Ukr.
J. Phys {\bf 41}, 370 (1996).
\bibitem{Tak79}
T. Takabayasi, Supp. Progr. Theor. Phys. {\bf 67}, 1 (1979);
W. N. Polyzou, Ann. Phys. (N.Y.) {\bf 193}, 367 (1989);
S. Ishida and M. Oda, Nuovo Cimento {\bf A107}, 2519 (1994).
\bibitem{HIK82}
M. Hirano, K. Iwata, K. Kato and T. Murota. Prog. Theor. Phys. {\bf 67},
1251 (1982);
D. D. Brayshaw, Phys. Rev. D {\bf 36}, 1465 (1987);
S. N. Grudtsin, IHEP Preprint 88-38, Serpukhov, 1988;
G. D. Tsibidis, hep-ph/0007143.
\bibitem{PPS87}
M. S. Plyushchay, G. P. Pron'ko, L. D. Soloviev, IHEP Preprint 87-24,
Serpukhov, 1987; V. I. Borodulin, M. S. Plyushchay and G. P. Pron'ko, Z.
Phys. C {\bf 41}, 293 (1988).
\bibitem{BNP92}
E. B. Berdnikov, G. P. Pron'ko, Sov. J. Nucl. Phys. {\bf 54}, 763 (1991);
E. B. Berdnikov, G. G. Nanobashvili, G. P. Pron'ko, Sov. J. Nucl. Phys.
{\bf 55}, 203 (1992); IHEP Preprint 92-67, Protvino, 1992.
\end{thebibliography}
